\documentclass[%
 reprint,
 amsmath,amssymb,
 aps,
]{revtex4-1}

\usepackage{graphicx}
\usepackage{dcolumn}
\usepackage{bm}

\begin{document}

\preprint{APS/123-QED}

\title{An insight of a needle free injector with magnetic material}

\author{Eun-Sung Jekal}
\affiliation{Department of material science, ETH Z\"urich, Z\"urich, Switzerland}%

\author{David Cooper, Samantha Maier, Seonyoung Park}
\affiliation{Harvard Medical School, Beth Israel Deaconess Medical Center, 330 Brookline Avenue, Boston, MA 02215, USA} 
\date{September 12, 2019}

\begin{abstract}
Needle-free injector technology (NFIT) has drawn attention due to their advantages. By using NFIT, it can be used several time unlike a conventional needle. Also, it  makes patients free from pain. Since NFIT by jet injection is achieved by ejecting a liquid drug through a narrow orifice at high pressure, thereby creating a fine high-speed fluid jet that can readily penetrate skin and tissue. Until very recently, all jet injectors utilized force-and pressure-generating principles that progress injection in anuncontrolled mannerwith limited ability to regulate delivery volume and injection depth. In order to address these shortcomings, we have developed a controllable jet injection  device,  based  on  a  custom  high-stroke linear  Lorentz-force motor. Using this device, we are able to monitor and modulate continuously the speed of the drug jet, and regulate  precisely  the  volume  of  drug  delivered  during  the  injection  process.
\end{abstract}

\maketitle


\section{Introduction}
Neeldle-free injection system is a novel method for the delivering of drug into patients \cite{mitragotri2006current,levine2003can,jackson2001safety,giudice2006needle}. Typically, drugs are delivered in the tissue by a disposable syringe. This manner of injection causes a lot of pain, and a needle is used only one time and be thrown out. Therefore, researchers have had attention to invent needle-free injection technology (NFIT). Among several. ways, Lorentz force plays a key role for push a pistion forward ejecting the drug. This would be possible at very high pressure and fast velocity. Moreover, the velocity is almost equal to that of sound in air \cite{taberner2012needle,hemond2006lorentz}.
 
For the Lorenze force actuator, magnetic material which surrounded by Cu coil and it facilitates the entire process. Unlike with conventional drug delivery systems, NFIT gives the user freedom from unnecessary pain and can be used multiple times. Furthermore, NFIT has shown promising results in mass immunization and vaccination programs.

NFIT requires strong energy to propel an premeasured dose of a particular drug formulation. Since these forces are generated from the ways ranging from high-pressure fluids including gases, electro-magnetic forces and shock waves to impart motion to the medicament, the NFIT device is built around a Lorentz force moving-coil actuator which consists of a magnet. This magnet is surrounded by a thin Cu wire coil that remains attached to a plastic piston which is inside a drug ampoule. A schematic diagram of present injector system using magnet is presented in Fig.1. In current research, a 3-layer coil surrounds an SmCo$_5$ magnet of 18 mm outer diameter. The coil is formed from 9.35 mm Cu wire wound onto a bobbin of 25 mm outer diameter. When current is applied, it interacts with the magnetic field so as to produce a force, which pushes the attached piston forward\cite{hogan2015needle,williams2012computational}.

\begin{figure}
    \centering
    \includegraphics[scale=0.5]{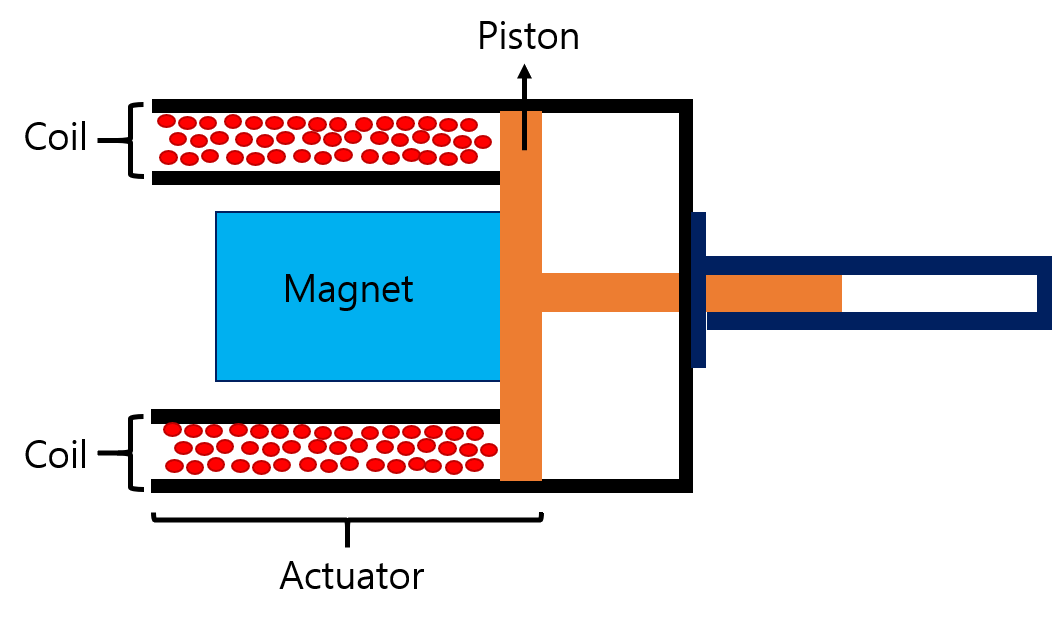}
    \caption{Schematic diagram of moving-coil injector system.}
    \label{fig:my_label}
\end{figure}

In case of the mechanical properties of the device and the fluid are modified, this model is used to investigate the behavior of an injector system, and thereby to derive the parameters that are most conducive to jet delivery of fluids\cite{chen2010needle,chang2013needle}.

The amount of current supplied can be very well regulated enabling the speed of the coil to come under our regulation. This would finally control the velocity with which the drug is ejected. 

\section{Results and discussions}
\subsection{Model formulation}
Since the elements of the injector form a coupled electo-mechanical and fluid system can be well descried by the block diagram, we present it in Fig.2. In this model, the voltage across the Cu coil ($V_C$) and it drives an electrical current. The current in the coil ($I_C$) is estimated using the empirically-measured impulse response function of the coil. A finite impulse response filter on the input voltage uses the impulse response function to calculate the current on a point-by-point basis within each simulation time step. The force produced by the coil ($F_C$) is calculated using $K_C$ as the force constant of the motor. The force constant is allowed to vary with coil position. The coil displacement ($x_C$) is estimated through a nonlinear mechanical model. The velocity of the coil induces a backEMF in the coil; the back-EMF is the source of a difference between the applied voltage ($V_A$) and the voltage across the coil.

The piston has a rubber tip that deforms to seal against the fluid when force is applied to the piston. The model combines the compliance of the tip and the piston shaft by ascribing a non-linear stiffness $k_P$ to the piston. The coil’s acceleration is described by 

\begin{equation}
\ddot{x}_c=\frac{F_c+k_P(x_P-x_C)}{m_c}.
\end{equation}

Here, we define $F_C$ is the force applied by the motor, while $x_C$ is the
displacement of the coil. $x_P$ is the displacement of the end of
the piston tip and $m_C$ is the mass of the coil. The piston tip acceleration is described by

\begin{equation}
\ddot{x}_P=\frac{-k_P(x_P-x_C)-F_{FR}-PA_P}{m_c}.
\end{equation}

$F_{FR}$ is the friction force and $P$ is the pressure of the fluid
within the ampoule, $A_P$ represents the area of the piston. Since the mass
of the piston is much less than the coil mass it is ignored.

Sliding friction arises at the interface between the rubber
piston tip and the walls of the ampoule and is proportional to
the pressure in the rubber piston tip, which is assumed to be
identical to the pressure in the fluid. Therefore, the friction
(F$_{FR}$) is calculated using

\begin{equation}
    F_{FR}=\mu A_CP,
\end{equation}

where $A_C$ is the contact area of the piston tip against the
ampoule wall and $\mu$ is the friction coefficient. A constant
level of static friction is present until the coil starts moving.
At the pressures encountered in jet injection, the compliance
of the fluid itself, measured by its bulk modulus, becomes
significant. In addition, the ampoule tends to expand under the
influence of the high pressure, contributing additional
compliance. The ampoule compliance is proportional to the
length (and thus the volume) of the fluid column, and can
therefore be lumped with the fluid compliance by way of an
effective ampoule-fluid bulk modulus ($K_{AF}$).

A differential equation for the change in pressure over time
can be determined via a mass balance. To account for the
effective bulk modulus, this equation was slightly modified to

\begin{equation}
    \dot{P}=\frac{(K_{AF}+P)\dot{x}_P-\frac{K_{AF}A_0}{A_P}u_0}{x_P}
\end{equation}

where $A_0$ is the area of the orifice and  is jet speed. The pressure that results from this equation was applied against the
piston as indicated in (2). The pressure loss due to viscous
fluid interactions is captured by

\begin{equation}
    P_{loss}=\frac{K_D}{2}\rho\mu_0 ^2,
\end{equation}

where $K_D$, the discharge coefficient, is empirically determined
and $\rho$ is the density of the fluid. $P_{loss}$ is subtracted from
pressure when calculating jet speed. Its formula can be
combined with Bernoulli’s equation and rearranged to incorporate viscous loss in (4),

\begin{equation}
    u_0=\sqrt{\frac{2P}{\rho(1+K_D)}}.
\end{equation}

The model was implemented in LabVIEW 2011 (National Instruments) using the Runge-Kutta-45 method within the Control and Simulation Module. This solved the model in 0.6 s.

\begin{figure}
    \centering
    \includegraphics[scale=0.5]{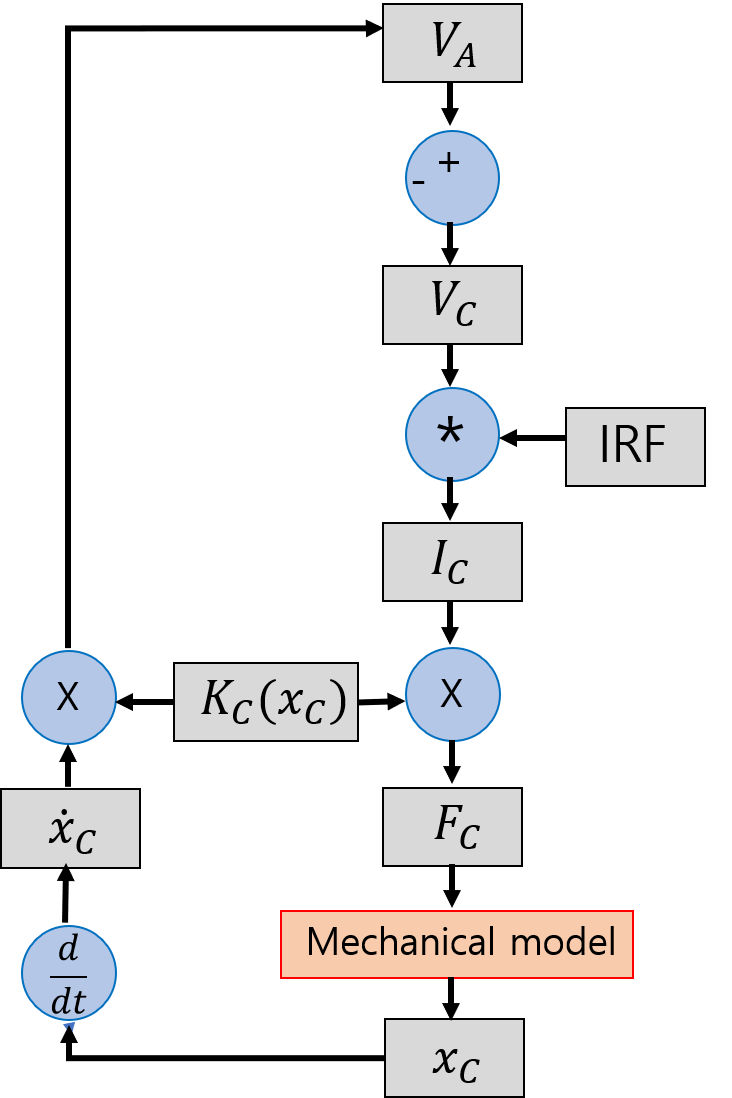}
    \caption{Block electro-mechanical diagram of moving-coil injector system.\cite{alford1995electro}}
    \label{fig:my_label}
\end{figure}

\subsection{Parameter Estimation}
We test the coil impedance impulse response function by using a low-noise linear amplifier. Gaussian white noise of ten volts peak amplitude was given to the coil. 
The sample rate of the input and output was about 100 kHz. The coil was fixed in position at 5 mm from full retraction. Using stochastic system identification, the measured voltage as well as current were applied to calculate the impulse response function. The impulse response function method produced model results that fit measured current values significantly better than a simple first-order 
series-circuit model of the inductance (4.8 mH) and resistance (9.4 $\omega$) of the coil.

For force production over stroke, a coupling mechanism was manufactured to connect the coil to a load cell. The force was measured at constant current over the stroke of the motor and a 2nd-order polynomial of force constant was fit to the results.
The average force constant over the stroke length of the injector was 8.78 N·A-1 and exhibited a standard deviation of 1.58 N·A-1. The maximum force constant measured was 10.2 N·A-1 and the minimum was 5.05 N·A-1.

For piston compliance, the compliance of the piston was
measured using an electromechanical test instrument. Epoxy was drawn into the ampoule up to the 0.05 mL mark and allowed to cure; the cured epoxy prevented the ampoule from compressing under load. Compression length.

The use of stochastic techniques to estimate the impulse response function allowed for the rise time of the modeled current to match measured current. As a result, the frequency of oscillations in modelled coil and piston tip position was well-aligned, providing a more accurate prediction of NFIT behavior.

By eliminating the effects of piston compliance, you can better predict the jet velocity profile. The model predicts a much smoother profile than the measured profile, but a slight mismatch may be due to noise.
The result is obvious vibrations in the coil position are not combined. Thus, we can conclude that when compliant pistons are used, they cannot rely on coil position to provide an acceptable jet velocity estimate during the dynamic phase of jet injection. Jet force measurements show a reasonable fit between force plate measurements and model prediction.

The problem is that vibration amplitude and vibration damping are not more accurately predicted than steady state jet velocity and jet velocity rise, due to the variability of the response to the compliant elements of the system.
Alternatively, the force transducer used may not have had the
necessary frequency response required to pick up the magnitude of the oscillations in the jet speed.
The model can be used to inform design decisions for future
development of jet injectors. As the device used in the
experiments is primarily used for fluids with viscosities near
to that of water, the model can be used to identify areas of
improvement for future viscous-drug devices. The effect of
piston compliance on the developed jet speed is negligible;
this is despite significant variation in the amplitude of
fluctuations for the coil position. Hence, a stiffer piston allows the potentiometer to better estimate the fluid volume and jet velocity, and tighter coupling between coil speed and jet
speed.

Obviously, overshoot must be reduced when we try to control jet speed. Ampoule suitability also plays an important role in determining. Rigid ampoules are easier to control injection due to the steady-state jet velocity, which is much faster than the original state.
When the injection speed is closely controlled, the depth between the jet speed and the injection depth can be identified. Thus, harder ampoules allow for smoother and more effective.

If the sliding friction coefficient is reduced, the jet velocity must increase without a change in the input energy from the coil. That is, the liquid must be constant, regardless of whether it is sticky or not. But the problem is how to reduce the sliding friction coefficient. Achieving this increases the efficiency of the system, making it easier to form high-speed jets, regardless of their viscosity.
The modeling will guide the future development such as drug type and wide use of the device. The first improvement we want to pursue is to strengthen the ampoules and pistons to improve the control and shape of the jet velocity profile over time. Secondly, we will utilize the control of the system to limit the rate of force applied to the coil to avoid excessive jet speed overshoot. If this is possible, finally the influence of the orifice diameter on the discharge coefficient and the investigation of the geometry of the ampoule will help to choose the best shape to force the fluid through.

\section{Conclusion}
In this paper we expand present knowledge odf NFIT. the NFTI electromechanical model. Via the characterization of the NFIT presented in this paper, we focused on identifying the elements of the system that most affect its performance. We identified these as the friction coefficient, the compliance of the NFIT components. The model can provides a better prediction of the jet speed profile over time than that provided by a direct conversion of coil speed to jet speed. For this perspective of view, the magnetic material is important. 

The insight that the model provides into jet development will be used to develop methods that improve the reliability of jet injection, focusing on delivering fluid to a particular layer underneath the skin surface with a targeted jet.

\bibliography{apssamp}

\begin{thebibliography}{11}%
\makeatletter
\providecommand \@ifxundefined [1]{%
 \@ifx{#1\undefined}
}%
\providecommand \@ifnum [1]{%
 \ifnum #1\expandafter \@firstoftwo
 \else \expandafter \@secondoftwo
 \fi
}%
\providecommand \@ifx [1]{%
 \ifx #1\expandafter \@firstoftwo
 \else \expandafter \@secondoftwo
 \fi
}%
\providecommand \natexlab [1]{#1}%
\providecommand \enquote  [1]{``#1''}%
\providecommand \bibnamefont  [1]{#1}%
\providecommand \bibfnamefont [1]{#1}%
\providecommand \citenamefont [1]{#1}%
\providecommand \href@noop [0]{\@secondoftwo}%
\providecommand \href [0]{\begingroup \@sanitize@url \@href}%
\providecommand \@href[1]{\@@startlink{#1}\@@href}%
\providecommand \@@href[1]{\endgroup#1\@@endlink}%
\providecommand \@sanitize@url [0]{\catcode `\\12\catcode `\$12\catcode
  `\&12\catcode `\#12\catcode `\^12\catcode `\_12\catcode `\%12\relax}%
\providecommand \@@startlink[1]{}%
\providecommand \@@endlink[0]{}%
\providecommand \url  [0]{\begingroup\@sanitize@url \@url }%
\providecommand \@url [1]{\endgroup\@href {#1}{\urlprefix }}%
\providecommand \urlprefix  [0]{URL }%
\providecommand \Eprint [0]{\href }%
\providecommand \doibase [0]{http://dx.doi.org/}%
\providecommand \selectlanguage [0]{\@gobble}%
\providecommand \bibinfo  [0]{\@secondoftwo}%
\providecommand \bibfield  [0]{\@secondoftwo}%
\providecommand \translation [1]{[#1]}%
\providecommand \BibitemOpen [0]{}%
\providecommand \bibitemStop [0]{}%
\providecommand \bibitemNoStop [0]{.\EOS\space}%
\providecommand \EOS [0]{\spacefactor3000\relax}%
\providecommand \BibitemShut  [1]{\csname bibitem#1\endcsname}%
\let\auto@bib@innerbib\@empty
\bibitem [{\citenamefont {Mitragotri}(2006)}]{mitragotri2006current}%
  \BibitemOpen
  \bibfield  {author} {\bibinfo {author} {\bibfnamefont {S.}~\bibnamefont
  {Mitragotri}},\ }\href@noop {} {\bibfield  {journal} {\bibinfo  {journal}
  {Nature reviews Drug discovery}\ }\textbf {\bibinfo {volume} {5}},\ \bibinfo
  {pages} {543} (\bibinfo {year} {2006})}\BibitemShut {NoStop}%
\bibitem [{\citenamefont {Levine}(2003)}]{levine2003can}%
  \BibitemOpen
  \bibfield  {author} {\bibinfo {author} {\bibfnamefont {M.~M.}\ \bibnamefont
  {Levine}},\ }\href@noop {} {\bibfield  {journal} {\bibinfo  {journal} {Nature
  medicine}\ }\textbf {\bibinfo {volume} {9}},\ \bibinfo {pages} {99} (\bibinfo
  {year} {2003})}\BibitemShut {NoStop}%
\bibitem [{\citenamefont {Jackson}\ \emph {et~al.}(2001)\citenamefont
  {Jackson}, \citenamefont {Austin}, \citenamefont {Chen}, \citenamefont
  {Stout}, \citenamefont {DeStefano}, \citenamefont {Gorse}, \citenamefont
  {Newman}, \citenamefont {Yu}, \citenamefont {Weniger}, \citenamefont {Group}
  \emph {et~al.}}]{jackson2001safety}%
  \BibitemOpen
  \bibfield  {author} {\bibinfo {author} {\bibfnamefont {L.~A.}\ \bibnamefont
  {Jackson}}, \bibinfo {author} {\bibfnamefont {G.}~\bibnamefont {Austin}},
  \bibinfo {author} {\bibfnamefont {R.~T.}\ \bibnamefont {Chen}}, \bibinfo
  {author} {\bibfnamefont {R.}~\bibnamefont {Stout}}, \bibinfo {author}
  {\bibfnamefont {F.}~\bibnamefont {DeStefano}}, \bibinfo {author}
  {\bibfnamefont {G.~J.}\ \bibnamefont {Gorse}}, \bibinfo {author}
  {\bibfnamefont {F.~K.}\ \bibnamefont {Newman}}, \bibinfo {author}
  {\bibfnamefont {O.}~\bibnamefont {Yu}}, \bibinfo {author} {\bibfnamefont
  {B.~G.}\ \bibnamefont {Weniger}}, \bibinfo {author} {\bibfnamefont {V.~S.
  D.~S.}\ \bibnamefont {Group}},  \emph {et~al.},\ }\href@noop {} {\bibfield
  {journal} {\bibinfo  {journal} {Vaccine}\ }\textbf {\bibinfo {volume} {19}},\
  \bibinfo {pages} {4703} (\bibinfo {year} {2001})}\BibitemShut {NoStop}%
\bibitem [{\citenamefont {Giudice}\ and\ \citenamefont
  {Campbell}(2006)}]{giudice2006needle}%
  \BibitemOpen
  \bibfield  {author} {\bibinfo {author} {\bibfnamefont {E.~L.}\ \bibnamefont
  {Giudice}}\ and\ \bibinfo {author} {\bibfnamefont {J.~D.}\ \bibnamefont
  {Campbell}},\ }\href@noop {} {\bibfield  {journal} {\bibinfo  {journal}
  {Advanced drug delivery reviews}\ }\textbf {\bibinfo {volume} {58}},\
  \bibinfo {pages} {68} (\bibinfo {year} {2006})}\BibitemShut {NoStop}%
\bibitem [{\citenamefont {Taberner}\ \emph {et~al.}(2012)\citenamefont
  {Taberner}, \citenamefont {Hogan},\ and\ \citenamefont
  {Hunter}}]{taberner2012needle}%
  \BibitemOpen
  \bibfield  {author} {\bibinfo {author} {\bibfnamefont {A.}~\bibnamefont
  {Taberner}}, \bibinfo {author} {\bibfnamefont {N.~C.}\ \bibnamefont {Hogan}},
  \ and\ \bibinfo {author} {\bibfnamefont {I.~W.}\ \bibnamefont {Hunter}},\
  }\href@noop {} {\bibfield  {journal} {\bibinfo  {journal} {Medical
  engineering \& physics}\ }\textbf {\bibinfo {volume} {34}},\ \bibinfo {pages}
  {1228} (\bibinfo {year} {2012})}\BibitemShut {NoStop}%
\bibitem [{\citenamefont {Hemond}\ \emph {et~al.}(2006)\citenamefont {Hemond},
  \citenamefont {Wendell}, \citenamefont {Hogan}, \citenamefont {Taberner},\
  and\ \citenamefont {Hunter}}]{hemond2006lorentz}%
  \BibitemOpen
  \bibfield  {author} {\bibinfo {author} {\bibfnamefont {B.~D.}\ \bibnamefont
  {Hemond}}, \bibinfo {author} {\bibfnamefont {D.~M.}\ \bibnamefont {Wendell}},
  \bibinfo {author} {\bibfnamefont {N.~C.}\ \bibnamefont {Hogan}}, \bibinfo
  {author} {\bibfnamefont {A.~J.}\ \bibnamefont {Taberner}}, \ and\ \bibinfo
  {author} {\bibfnamefont {I.~W.}\ \bibnamefont {Hunter}},\ }in\ \href@noop {}
  {\emph {\bibinfo {booktitle} {2006 International Conference of the IEEE
  Engineering in Medicine and Biology Society}}}\ (\bibinfo {organization}
  {IEEE},\ \bibinfo {year} {2006})\ pp.\ \bibinfo {pages}
  {679--682}\BibitemShut {NoStop}%
\bibitem [{\citenamefont {Hogan}\ \emph {et~al.}(2015)\citenamefont {Hogan},
  \citenamefont {Taberner}, \citenamefont {Jones},\ and\ \citenamefont
  {Hunter}}]{hogan2015needle}%
  \BibitemOpen
  \bibfield  {author} {\bibinfo {author} {\bibfnamefont {N.~C.}\ \bibnamefont
  {Hogan}}, \bibinfo {author} {\bibfnamefont {A.~J.}\ \bibnamefont {Taberner}},
  \bibinfo {author} {\bibfnamefont {L.~A.}\ \bibnamefont {Jones}}, \ and\
  \bibinfo {author} {\bibfnamefont {I.~W.}\ \bibnamefont {Hunter}},\
  }\href@noop {} {\bibfield  {journal} {\bibinfo  {journal} {Expert opinion on
  drug delivery}\ }\textbf {\bibinfo {volume} {12}},\ \bibinfo {pages} {1637}
  (\bibinfo {year} {2015})}\BibitemShut {NoStop}%
\bibitem [{\citenamefont {Williams}\ \emph {et~al.}(2012)\citenamefont
  {Williams}, \citenamefont {Hogan}, \citenamefont {Nielsen}, \citenamefont
  {Hunter},\ and\ \citenamefont {Taberner}}]{williams2012computational}%
  \BibitemOpen
  \bibfield  {author} {\bibinfo {author} {\bibfnamefont {R.~M.}\ \bibnamefont
  {Williams}}, \bibinfo {author} {\bibfnamefont {N.~C.}\ \bibnamefont {Hogan}},
  \bibinfo {author} {\bibfnamefont {P.~M.}\ \bibnamefont {Nielsen}}, \bibinfo
  {author} {\bibfnamefont {I.~W.}\ \bibnamefont {Hunter}}, \ and\ \bibinfo
  {author} {\bibfnamefont {A.~J.}\ \bibnamefont {Taberner}},\ }in\ \href@noop
  {} {\emph {\bibinfo {booktitle} {2012 Annual International Conference of the
  IEEE Engineering in Medicine and Biology Society}}}\ (\bibinfo {organization}
  {IEEE},\ \bibinfo {year} {2012})\ pp.\ \bibinfo {pages}
  {2052--2055}\BibitemShut {NoStop}%
\bibitem [{\citenamefont {Chen}\ \emph {et~al.}(2010)\citenamefont {Chen},
  \citenamefont {Lv},\ and\ \citenamefont {Zhou}}]{chen2010needle}%
  \BibitemOpen
  \bibfield  {author} {\bibinfo {author} {\bibfnamefont {K.}~\bibnamefont
  {Chen}}, \bibinfo {author} {\bibfnamefont {Y.}~\bibnamefont {Lv}}, \ and\
  \bibinfo {author} {\bibfnamefont {H.}~\bibnamefont {Zhou}},\ }in\ \href@noop
  {} {\emph {\bibinfo {booktitle} {2010 International Conference on Mechanic
  Automation and Control Engineering}}}\ (\bibinfo {organization} {IEEE},\
  \bibinfo {year} {2010})\ pp.\ \bibinfo {pages} {5310--5314}\BibitemShut
  {NoStop}%
\bibitem [{\citenamefont {Chang}\ \emph {et~al.}(2013)\citenamefont {Chang},
  \citenamefont {Hogan},\ and\ \citenamefont {Hunter}}]{chang2013needle}%
  \BibitemOpen
  \bibfield  {author} {\bibinfo {author} {\bibfnamefont {J.~H.}\ \bibnamefont
  {Chang}}, \bibinfo {author} {\bibfnamefont {N.~C.}\ \bibnamefont {Hogan}}, \
  and\ \bibinfo {author} {\bibfnamefont {I.~W.}\ \bibnamefont {Hunter}},\ }in\
  \href@noop {} {\emph {\bibinfo {booktitle} {2013 35th Annual International
  Conference of the IEEE Engineering in Medicine and Biology Society (EMBC)}}}\
  (\bibinfo {organization} {IEEE},\ \bibinfo {year} {2013})\ pp.\ \bibinfo
  {pages} {3491--3494}\BibitemShut {NoStop}%
\bibitem [{\citenamefont {Alford}\ and\ \citenamefont
  {Rupert}(1995)}]{alford1995electro}%
  \BibitemOpen
  \bibfield  {author} {\bibinfo {author} {\bibfnamefont {R.~L.}\ \bibnamefont
  {Alford}}\ and\ \bibinfo {author} {\bibfnamefont {J.~G.}\ \bibnamefont
  {Rupert}},\ }\href@noop {} {\enquote {\bibinfo {title} {Electro-mechanical
  roll control apparatus and method},}\ } (\bibinfo {year} {1995}),\ \bibinfo
  {note} {uS Patent 5,452,864}\BibitemShut {NoStop}%
\end{thebibliography}%

\end{document}